\documentclass[12pt]{article}
\usepackage{epsf}
\begin{document}

\newcommand{\gtrsim}{ \mathop{}_{\textstyle \sim}^{\textstyle >} }
\newcommand{\lesssim}{ \mathop{}_{\textstyle \sim}^{\textstyle <} }

\newcommand{\rem}[1]{{\bf #1}}

\renewcommand{\thefootnote}{\fnsymbol{footnote}}
\setcounter{footnote}{0}
\begin{titlepage}

\def\thefootnote{\fnsymbol{footnote}}

\begin{center}

\hfill IASSNS-HEP-00/12\\
\hfill hep-ph/0002208\\
\hfill February, 2000\\

\vskip .75in

{\Large \bf
  Effects of the right-handed neutrinos on $\Delta S=2$ and 
  $\Delta B=2$ processes in supersymmetric SU(5) model
}

\vskip .75in

{\large
  Takeo Moroi
}

\vskip 0.25in

{\em
  School of Natural Sciences, Institute for Advanced Study\\
  Olden Lane, Princeton, NJ 08540, U.S.A.}

\end{center}

\vskip .5in

\begin{abstract}

We discuss an extra source of CP and flavor violations in
supersymmetric SU(5) grand unified model with the right-handed
neutrinos.  In such a model, the right-handed down-type squarks
$\tilde{d}_R$ interact with the right-handed neutrinos above the GUT
scale, and the renormalization group effect can generate sizable
off-diagonal elements in the mass matrix of $\tilde{d}_R$.  Because of
new Yukawa phases which exist in the SU(5) model, these off-diagonal
elements have, in general, large CP violating phases.  The
renormalization group induced off-diagonal elements affect the $K$ and
$B$ decays.  In particular, in this model, supersymmetric contribution
to the $\epsilon_K$ parameter can be as large as the currently
measured experimental value, and hence the effect might be seen as an
anomaly in the on-going test of the Cabibbo-Kobayashi-Maskawa
triangle.

\end{abstract}
\end{titlepage}

\renewcommand{\thepage}{\arabic{page}}
\setcounter{page}{1}
\renewcommand{\thefootnote}{\#\arabic{footnote}}
\setcounter{footnote}{0}

In particle physics, one of the greatest excitement in the last decade
is the discovery of the evidence for the neutrino
oscillations~\cite{SuzTot}.  In particular, the anomalies in the
atmospheric and solar neutrino fluxes suggest neutrino masses much
smaller than those of the quarks and charged leptons.

Such small neutrino masses are beautifully explained by the see-saw
mechanism with the right-handed neutrinos~\cite{Yanagida, GelRamSla}.
The see-saw mechanism gives the (left-handed) neutrino masses of the
form $m_{\nu_L}\sim m_{\nu_{\rm D}}^2/M_{\nu_R}$, where $m_{\nu_{\rm
D}}$ and $M_{\nu_R}$ are Dirac and Majorana masses for neutrinos,
respectively.  The Dirac mass $m_{\nu_{\rm D}}$ is from the Yukawa
interaction with the electroweak Higgs boson, and is of order the
electroweak scale $M_{\rm weak}$ (or smaller).  The Majorana mass
$M_{\nu_R}$ is, however, not related to the electroweak symmetry
breaking and can be much larger than $M_{\rm weak}$.  Adopting
$M_{\nu_R}\gg M_{\rm weak}$, $m_{\nu_L}\ll M_{\rm weak}$ is realized.
For example, assuming $m_{\nu_{\rm D}}\sim M_{\rm weak}$, the
atmospheric neutrino anomaly predicts $M_{\nu_R}\sim 10^{14} -
10^{15}~{\rm GeV}$ with relevant mixing parameters.

This fact suggests the validity of the field theoretical description
up to a scale much higher than the electroweak scale.  If so, the
quadratic divergence in the Higgs boson mass parameter badly
destabilize the electroweak scale unless some mechanism protects the
stability.  A natural solution to this problem is to introduce
supersymmetry (SUSY); in SUSY models, quadratic divergences are
cancelled between bosonic and fermionic loops and hence the
electroweak scale becomes stable against radiative corrections.  SUSY
standard model also suggests an interesting new physics at a high
scale, i.e., the grand unified theory (GUT).  With the renormalization
group (RG) equation based on the minimal SUSY standard model (MSSM),
all the gauge coupling constants meet at $M_{\rm GUT}\simeq 2\times
10^{16}~{\rm GeV}$, while the gauge coupling unification is not
realized in the (non-supersymmetric) standard model \cite{PRD44-817,
PLB260-447}.

Therefore, SUSY GUT with the right-handed neutrinos is a
well-motivated extension of the standard model.  In this paper, we
consider $\Delta S=2$ and $\Delta B=2$ processes in the SU(5) GUT with
the right-handed neutrinos.  Before the Super-Kamiokande experiment
\cite{PRL88-1562}, similar effects were considered without taking into
account the large mixing in the neutrino sector \cite{NPB449-437}.
Now, we have better insights into the neutrino mass and mixing
parameters.  In particular, the atmospheric neutrino flux measured by
the Super-Kamiokande experiment strongly favors a large 2-3 mixing in
the neutrino sector \cite{PRL88-1562}.  Furthermore, with the on-going
$B$ factories, the mixings in the quark sector, in particular, the
Cabibbo-Kobayashi-Maskawa (CKM) matrix elements, will be well
constrained.  In such a circumstance, it is appropriate to study
implications of the right-handed neutrinos to the $\Delta S=2$ and
$\Delta B=2$ processes within the current experimental situation.

In grand unified models, the right-handed down type squarks
$\tilde{d}_R$ couple to the right-handed neutrinos above the GUT
scale, and sizable flavor violations are possible in the $\tilde{d}_R$
sector.  In particular, the RG effect may generate off-diagonal
elements in the soft SUSY breaking mass matrix of $\tilde{d}_R$.
(Notice that the flavor violation in the $\tilde{d}_R$ sector is
highly suppressed in the minimal supergravity model without the
right-handed neutrinos.)  Furthermore, in grand unified models, there
are new physical CP violating phases in the colored Higgs vertices,
which do not affect the low energy Yukawa interactions.  These may
affect CP and flavor violating processes in $K$ and $B$ system.
Indeed, as we will see, SUSY contribution to the $\epsilon_K$
parameter can be as large as the experimental value even with a
relatively heavy squark mass $m_{\tilde{q}}\sim 1~{\rm TeV}$, and it
might cause an anomaly in the on-going test of the CKM triangle.

We start by introducing the Lagrangian for the SUSY SU(5) model
with the right-handed neutrinos.  Denoting ${\bf 10}$, ${\bf
\bar{5}}$, and singlet fields of SU(5) as $\Psi_i$, $\Phi_i$, and
$N_i$, respectively, the superpotential is given by\footnote{We assume
that other Yukawa couplings (in particular, those of the SU(5)
symmetry breaking fields) are small.  We neglect their effects in this
paper.}
\begin{eqnarray}
W_{\rm GUT} = \frac{1}{8} \Psi_i \left[Y_{U}\right]_{ij} \Psi_j H
+ \Psi_i \left[Y_{D}\right]_{ij} \Phi_j \bar{H} 
+ N_i \left[Y_{N}\right]_{ij} \Phi_j H
+ \frac{1}{2} N_i \left[M_{N}\right]_{ij} N_j,
\label{W_GUT}
\end{eqnarray}
where $\bar{H}$ and $H$ are Higgs fields which are in ${\bf \bar{5}}$
and ${\bf 5}$ representations, respectively, and $i$ and $j$ are
generation indices which run from 1 to 3.  (We omit the SU(5) indices
for the simplicity of the notation.)  Here, $M_N$ is the Majorana mass
matrix of the right-handed neutrinos.  In order to make our points
clearer, we adopt the simplest Majorana mass matrix, that is, the
universal structure of $M_N$:
\begin{eqnarray}
\left[M_{N}\right]_{ij} = M_{\nu_R} \delta_{ij}.
\label{M_nR}
\end{eqnarray}
The following discussions are qualitatively unaffected by this
assumption, although numerical results depend on the structure of
$M_N$.  In Eq.~(\ref{W_GUT}), $Y$'s are $3\times 3$ Yukawa matrices;
in general, $Y_{U}$ is a general complex symmetric matrix, while
$Y_{D}$ and $Y_{N}$ are general complex matrices.  By choosing a
particular basis, we can eliminate unphysical phases and angles.  In
this paper, we choose the basis of $\Psi_i$, $\Phi_i$, and $N_i$ such
that the Yukawa matrices at the GUT scale become
\begin{eqnarray}
Y_U (M_{\rm GUT}) = V_Q^T \hat{\Theta}_Q \hat{Y}_U V_Q,~~~
Y_D (M_{\rm GUT}) = \hat{Y}_D,~~~
Y_N (M_{\rm GUT}) = \hat{Y}_N V_L \hat{\Theta}_L,
\end{eqnarray}
where $\hat{Y}$'s are real diagonal matrices:
\begin{eqnarray}
\hat{Y}_U = {\rm diag} (y_{u_1},  y_{u_2},  y_{u_3}),~~~
\hat{Y}_D = {\rm diag} (y_{d_1},  y_{d_2},  y_{d_3}),~~~
\hat{Y}_N = {\rm diag} (y_{\nu_1},  y_{\nu_2},  y_{\nu_3}),
\end{eqnarray}
while $\hat{\Theta}$'s are diagonal phase matrices:
\begin{eqnarray}
\hat{\Theta}_Q = 
{\rm diag}
(e^{i\phi^{(Q)}_1}, e^{i\phi^{(Q)}_2}, e^{i\phi^{(Q)}_3}),~~~
\hat{\Theta}_L = 
{\rm diag}
(e^{i\phi^{(L)}_1}, e^{i\phi^{(L)}_2}, e^{i\phi^{(L)}_3}),
\end{eqnarray}
where phases obey the constraints
$\phi^{(Q)}_1+\phi^{(Q)}_2+\phi^{(Q)}_3=0$ and
$\phi^{(L)}_1+\phi^{(L)}_2+\phi^{(L)}_3=0$.  Furthermore, $V_Q$ and
$V_L$ are unitary mixing matrices and are parameterized as \cite{PDG}
\begin{eqnarray}
V_Q = 
\left(
\begin{array}{ccc}
 c^{(Q)}_{12} c^{(Q)}_{13} & 
 s^{(Q)}_{12} c^{(Q)}_{13} & 
 s^{(Q)}_{13} e^{-i\delta_{13}^{(Q)}} \\
-s^{(Q)}_{12} c^{(Q)}_{23} 
-c^{(Q)}_{12} s^{(Q)}_{23} s^{(Q)}_{13} e^{i\delta_{13}^{(Q)}} &
 c^{(Q)}_{12} c^{(Q)}_{23} 
-s^{(Q)}_{12} s^{(Q)}_{23} s^{(Q)}_{13} e^{i\delta_{13}^{(Q)}} &
 s^{(Q)}_{23} c^{(Q)}_{13} \\
 s^{(Q)}_{12} s^{(Q)}_{23}
-c^{(Q)}_{12} c^{(Q)}_{23} s^{(Q)}_{13} e^{i\delta_{13}^{(Q)}} &
-c^{(Q)}_{12} s^{(Q)}_{23} 
-s^{(Q)}_{12} c^{(Q)}_{23} s^{(Q)}_{13} e^{i\delta_{13}^{(Q)}} &
 c^{(Q)}_{23} c^{(Q)}_{13}
\end{array}
\right),
\end{eqnarray}
where $c^{(Q)}_{ij}=\cos\theta^{(Q)}_{ij}$ and
$s^{(Q)}_{ij}=\sin\theta^{(Q)}_{ij}$, and $V_L = V_Q|_{Q\rightarrow
L}$.

$W_{\rm GUT}$ can be also expressed by using the standard model
fields.  By properly relating the GUT fields with the standard model
fields, the phases $\phi^{(Q)}$ and $\phi^{(L)}$ are eliminated from
the interactions among the light fields.  Indeed, let us embed the
standard model fields into the GUT fields as
\begin{eqnarray}
\Psi_i \simeq
\{ 
Q,\ V_Q^\dagger \hat{\Theta}_Q^\dagger \bar{U},\ 
\hat{\Theta}_L \bar{E} 
\}_i,~~~
\Phi_i \simeq \{ \bar{D},\ \hat{\Theta}_L^\dagger L \}_i,
\end{eqnarray}
where $Q_i ({\bf 3},{\bf 2})_{1/6}$, $\bar{U}_i ({\bf \bar{3}},{\bf
1})_{-2/3}$, $\bar{D}_i ({\bf \bar{3}},{\bf 1})_{1/3}$, $L_i ({\bf
1},{\bf 2})_{1/2}$, and $\bar{E}_i ({\bf 1},{\bf 1})_{1}$ are quarks
and leptons in $i$-th generation with the SU(3)$_C$ $\times$ SU(2)$_L$
$\times$ U(1)$_Y$ gauge quantum numbers as indicated.  With this
embedding, $W_{\rm GUT}$ at the GUT scale becomes
\begin{eqnarray}
W_{\rm GUT} &=& W_{\rm SSM}
\nonumber \\ &&
- \frac{1}{2} Q_i
\left[V_Q^T \hat{\Theta}_Q \hat{Y}_U V_Q \right]_{ij} Q_j H_C
+ \bar{E}_i \left[ \hat{\Theta}_L \hat{Y}_U \right]_{ij} \bar{U}_j H_C
\nonumber \\ &&
+ \bar{U}_i \left[ \hat{\Theta}_Q^\dagger V_Q^* \hat{Y}_D \right]_{ij}
\bar{D}_j \bar{H}_C
- Q_i \left[ \hat{Y}_D \hat{\Theta}_L^\dagger\right]_{ij} 
L_j \bar{H}_C
\nonumber \\ &&
+ N_i \left[ \hat{Y}_N V_L \hat{\Theta}_L \right]_{ij} \bar{D}_j H_C,
\label{W_Hc}
\end{eqnarray}
where $H_C$ and $\bar{H}_C$ are colored Higgses, and the low energy
Lagrangian is given by
\begin{eqnarray}
W_{\rm SSM} &=&
Q_i \left[ V_Q^T \hat{Y}_U \right]_{ij} \bar{U}_j H_u
+ Q_i \left[ \hat{Y}_D \right]_{ij} \bar{D}_j H_d
+ \bar{E}_i \left[ \hat{Y}_E \right]_{ij} L_j H_d
+ N_i \left[ \hat{Y}_N V_L \right]_{ij} L_j H_u
\nonumber \\ &&
+ \frac{1}{2} M_{\nu_R} N_i N_i.
\label{W_SSM}
\end{eqnarray}
with $H_u$ and $H_d$ being the up- and down-type Higgses,
respectively.  Simple SU(5) GUT predicts $\hat{Y}_D=\hat{Y}_E$, but
this relation breaks down for the first and second generations.  Some
non-trivial flavor physics is necessary to have a realistic down-type
quark and charged lepton Yukawa matrices.  Such a new physics may
provide a new source of flavor and CP violations \cite{PRD53-413,
PRD58-116010}, but we do not consider such an effect in this paper.

In the basis given in Eq.~(\ref{W_SSM}), Yukawa matrices for the
down-type quarks and charged leptons are diagonal.  Strictly speaking,
the RG effect mixes the flavors and their diagonalities are not
preserved at lower scale.  Such a flavor mixing is, however, rather a
minor effect for the Yukawa matrices and the down-type quarks and
charged leptons in Eq.~(\ref{W_SSM}) correspond to the mass
eigenstates quite accurately.  Therefore, this basis is useful to
understand the qualitative features of the $K$ and $B$ physics.  So,
we first discuss the flavor violation of the model using
Eq.~(\ref{W_SSM}) neglecting the RG effect for the Yukawa matrices
below the GUT scale.\footnote{Notice that, in our numerical discussion
later, all the flavor mixing effects are included.  In particular,
mass eigenstates of the quarks and charged leptons will be defined
using the Yukawa matrices given at the electroweak scale.}

With the superpotential (\ref{W_SSM}), the left-handed neutrino mass
matrix is given by
\begin{eqnarray}
[m_{\nu_L}]_{ij} = \frac{\langle H_u\rangle^2}{M_{\nu_R}} 
\left[ V_L^T \hat{Y}_N^2 V_L \right]_{ij}
= \frac{v^2\sin^2\beta}{2M_{\nu_R}} 
\left[ V_L^T \hat{Y}_N^2 V_L \right]_{ij},
\label{nuL_mass}
\end{eqnarray}
where $v\simeq 246~{\rm GeV}$, and $\beta$ parameterizes the relative
size of two Higgs vacuum expectation values: $\tan\beta\equiv\langle
H_u\rangle /\langle H_d\rangle$.  Then, $V_L$ provides the mixing
matrix of the lepton sector.  On the contrary, as seen in
Eq.~(\ref{W_SSM}), $V_Q$ is the CKM matrix: $V_Q\simeq V_{\rm CKM}$.

At the tree level, effects of the heavy particles, like the colored
Higgses and right-handed neutrinos, are suppressed by inverse powers
of their masses.  As a result, we can mostly completely neglect the
interaction with such heavy particles, although there are a few
important processes like the neutrino oscillation and the proton
decay.  The phases $\phi^{(Q)}$ and $\phi^{(L)}$ do not affect the
Yukawa interactions among the standard model fields and only appear in
the colored Higgs vertices.  Therefore, their effects on low energy
physics are rather indirect, and those phases are not determined from
the Yukawa interactions of the quarks and leptons with the electroweak
Higgses.

At the loop level, however, heavy particles may affect the low energy
quantities through the RG effect
\cite{NPB267-415,PRL57-961,PLB321-56,PLB338-212}.  In particular,
flavor changing operators may be significantly affected.  For example,
the neutrino Yukawa interactions may induce off-diagonal elements in
the soft SUSY breaking mass matrix of left-handed sleptons.  In
addition, even if there is no neutrino Yukawa interactions, lepton
flavor numbers are violated in the grand unified models by the up-type
Yukawa interaction, and hence we may expect non-vanishing off-diagonal
elements in the mass matrix of the right-handed sleptons.  Since the
lepton flavor numbers are preserved in the standard model, these
effects result in very drastic signals like $\mu\rightarrow e\gamma$
and $\tau\rightarrow\mu\gamma$ in SU(5) \cite{PLB338-212, NPB445-219,
PLB391-341} and SO(10) \cite{NPB445-219, NPB458-3, PRD53-413} models
as well as in models with the right-handed neutrinos \cite{PRL57-961,
PLB357-579, PRD53-2442, PRD59-116005}.  Furthermore, hadronic flavor
and CP violations in grand unified models were considered in
Ref.~\cite{NPB449-437} without taking into account the large mixing in
the neutrino sector.  The flavor mixing below the GUT scale through
the CKM matrix also affects the $\Delta S=2$ and $\Delta B=2$
processes \cite{PRD39-3447, NPB353-591, PRD53-5233, hph9908449}.
Here, we consider CP violation and flavor mixing in the down sector
induced by the neutrino Yukawa matrix.  As seen in Eq.~(\ref{W_Hc}),
$\bar{D}$ interacts with the right-handed neutrinos above the GUT
scale, and its soft SUSY breaking scalar mass matrix is affected by
the neutrino Yukawa interactions.  Because of possible flavor and CP
violating parameters in $V_L$ and $\hat{\Theta}_L$, this can be
important.  Similar effect was also considered in
Ref.~\cite{PRD59-116005} for the $b\rightarrow s\gamma$ process, for
which the effect was insignificant.  Here, we will show that a large
effect is possible in particular in the CP violation in the $K$ decay.
Furthermore, Ref.~\cite{PRD59-116005} did not take into account the
physical phases $\phi^{(L)}$ which play a very important role in
calculating the SUSY contribution to the $\epsilon_K$ parameter.

Now, we are at the point to discuss the RG effect and to estimate the
off-diagonal elements in the down-type squark mass matrix.  The RG
effect is not calculable unless the boundary condition of the SUSY
breaking parameters are given.  In order to estimate the importance of
this effect, we use the minimal supergravity boundary condition.
Then, the SUSY breaking parameters are parameterized by the following
three parameters:\footnote{There are extra important parameters.  In
particular, in order to have a viable electroweak symmetry breaking,
so-called $\mu$- and $B_\mu$-parameters are necessary.  However, they
do not affect the running of other parameters, so they do not have to
be specified in our RG analysis.  We assume they have correct values
required from the radiative electroweak breaking.}  the universal
scalar mass $m_0$, the universal $A$-parameter $a_0$ which is the
ratio of SUSY breaking trilinear scalar interactions to corresponding
Yukawa couplings, and the SU(5) gaugino mass $m_{G5}$.  We assume that
the boundary condition is given at the reduced Planck scale $M_*\simeq
2.4\times 10^{18}~{\rm GeV}$.

Once the boundary condition is given, the MSSM Lagrangian at the
electroweak scale is obtained by running all the parameters down from
the reduced Planck scale to the electroweak scale with RG equation.
For the RG analysis, relevant effective field theory describing the
energy scale $\mu$ is MSSM (without the right-handed neutrinos) for
$M_{\rm weak}\le\mu\le M_{\nu_R}$, SUSY standard model with the
right-handed neutrinos for $M_{\nu_R}\le\mu\le M_{\rm GUT}$, and SUSY
SU(5) GUT for $M_{\rm GUT}\le\mu\le M_*$.  We use the minimal SUSY
SU(5) GUT where the SU(5) symmetry is broken by {\bf 24}
representation of SU(5).

Before showing the results of the numerical calculation, let us
discuss the behavior of the solution to the RG equation using a simple
approximation. With the one-iteration approximation, off-diagonal
elements in the soft SUSY breaking mass of $\tilde{d}_R$ is given by
\begin{eqnarray}
[m^2_{\tilde{d}_R}]_{ij} &\simeq&
-\frac{1}{8\pi^2} \left[ Y_N^\dagger Y_N \right]_{ij} (3m_0^2+a_0^2)
\log \frac{M_*}{M_{\rm GUT}}
\nonumber \\ &\simeq&
-\frac{1}{8\pi^2} e^{-i(\phi^{(L)}_i-\phi^{(L)}_j)}
y_{\nu_k}^2 \left[V_{L}^*\right]_{ki} \left[V_{L}\right]_{kj}
(3m_0^2+a_0^2) \log \frac{M_*}{M_{\rm GUT}}.
\label{m_dR}
\end{eqnarray}
This expression suggests two important consequences of the SU(5) model
with the right-handed neutrinos.  First, if neutrinos have a large
Yukawa coupling and mixing, off-diagonal elements in
$[m^2_{\tilde{d}_R}]_{ij}$ are generated.  Since the atmospheric and
solar neutrino anomalies require non-vanishing mixing parameters in
$V_L$, this effect may generate sizable off-diagonal elements in
$[m^2_{\tilde{d}_R}]$.  Notice that, below the GUT scale,
$\tilde{d}_R$ have very weak flavor violating Yukawa interactions.  As
a result, if the minimal supergravity boundary condition is given at
the GUT scale, we see much smaller flavor violation in
$[m^2_{\tilde{d}_R}]$.  The second point is on the phase of the RG
induced off-diagonal elements.  Because the phases $\phi^{(L)}$ are
free parameters, the RG induced off-diagonal elements (\ref{m_dR}) may
have arbitrary phases and large CP violating phases in the mass matrix
of $\tilde{d}_R$ are possible.

As can be understood in Eq.~(\ref{m_dR}), $[m^2_{\tilde{d}_R}]_{ij}$
depends on the structure of the neutrino mixing matrix $V_L$.  In our
analysis, we use the mixing angles and neutrino masses suggested by
the atmospheric neutrino data, and also by the large angle or small
angle MSW solution to the solar neutrino problem
\cite{hph9802201}.\footnote{We can also consider the case of the
vacuum oscillation solution to the solar neutrino problem.  The vacuum
oscillation solution requires very light second generation neutrino
($m_{\nu_2}\sim O(10^{-5}~{\rm eV})$).  With our simple set up,
$y_{\nu_2}$ cannot be large once the perturbativity of $y_{\nu_3}$ is
imposed up to $M_*$, since we assume the universal structure of $M_N$.
Consequently, the RG effect is suppressed except for the 2-3 mixing.
If we consider general neutrino mass matrix, however, this may not be
true.  For example, non-universal $M_N$ or non-vanishing
$[V_{L}]_{31}$ would easily enhance the RG effect.}  For the large
angle MSW case, we use
\begin{eqnarray}
m_{\nu} \simeq
( \begin{array}{ccc}
0, & 0.004~{\rm eV}, & 0.03~{\rm eV}
\end{array} ),~~~
V_L \simeq
\left( \begin{array}{ccc}
0.91 & -0.30 & 0.30 \\
0.42 & 0.64 & -0.64 \\
0 & 0.71 & 0.70
\end{array} \right),
\label{VL_la}
\end{eqnarray}
and for the small angle case, we use
\begin{eqnarray}
m_{\nu} \simeq 
( \begin{array}{ccc} 
0, & 0.003~{\rm eV}, & 0.03~{\rm eV}
\end{array} ),~~~
V_L \simeq
\left( \begin{array}{ccc}
1.0 & -0.028 & 0.028 \\
0.040 & 0.70 & -0.71 \\
0 & 0.71 & 0.70
\end{array} \right).
\label{VL_sa}
\end{eqnarray}
Here, we assume that the neutrino masses are hierarchical, and that
the lightest neutrino mass is negligibly small.  The 3-1 element of
$V_L$ is known to be relatively small: $|[V_L]_{31}|^2\leq 0.05$
\cite{hph9802201} from the CHOOZ experiment \cite{PLB420-397}.  We
simply assume $[V_L]_{31}$ be small enough to be neglected, and we
take $[V_L]_{31}=0$.  We will discuss the implication of
$[V_L]_{31}\neq 0$ later.  In calculating the neutrino Yukawa matrix,
we take $M_{\nu_R}$ as a free parameter, and the neutrino Yukawa
matrix is determined as a function of $M_{\nu_R}$ using
Eq.~(\ref{nuL_mass}).

In order to discuss the flavor violating effects, it is convenient to
define $\Delta^{(R)}_{ij}$, which is the off-diagonal elements in
$[m^2_{\tilde{d}_R}]_{ij}$ normalized by the squark mass:
\begin{eqnarray}
\Delta^{(R)}_{ij}\equiv 
\frac{[m^2_{\tilde{d}_R}]_{ij}}{m^2_{\tilde{q}}} \equiv
\frac{[m^2_{\tilde{d}_R}]_{ij}}{[m^2_{\tilde{d}_R}]_{11}},
\label{delta_ij}
\end{eqnarray}
where all the quantities are evaluated at the electroweak scale.
Notice that, in our case, all the squark masses are almost degenerate
except those for the stops. We use the first generation right-handed
squark mass as a representative value.

\begin{figure}
\centerline{\epsfxsize=0.5\textwidth\epsfbox{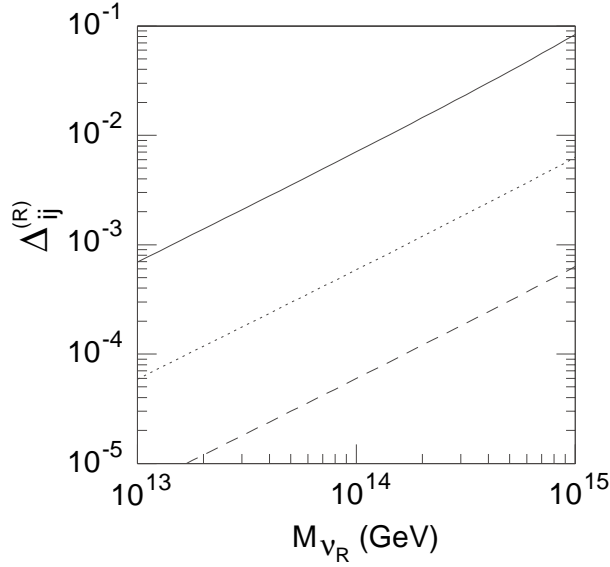}}
\caption{The flavor violation parameters defined in
Eq.~(\ref{delta_ij}); $|\Delta^{(R)}_{23}|$ in the large angle MSW
case (solid), $|\Delta^{(R)}_{12}|$ in the large angle MSW case
(dotted), and $|\Delta^{(R)}_{12}|$ in the small angle MSW case
(dashed).  We take $a_0=0$, $\tan\beta =3$, and $m_{G5}$ and $m_0$ are
chosen so that $m_{\tilde{q}}=1~{\rm TeV}$ and $m_{\tilde{W}}=150~{\rm
GeV}$.}
\label{fig:dij}
\end{figure}

In Fig.~\ref{fig:dij}, we plot several $|\Delta^{(R)}_{ij}|$ as a
function of $M_{\nu_R}$. Here, we take $a_0=0$, $\tan\beta =3$, and
$m_{G5}$ and $m_0$ are chosen so that $m_{\tilde{q}}=1~{\rm TeV}$ and
$m_{\tilde{W}}=150~{\rm GeV}$ (which gives the gluino mass of
$m_{\tilde{G}}\simeq 520~{\rm GeV}$).  As one can see, for larger
$M_{\nu_R}$, off-diagonal elements are more enhanced.  This is because
the neutrino Yukawa coupling is proportional to $M_{\nu_R}^{1/2}$ for
fixed neutrino mass.  $\Delta^{(R)}_{23}$ is dominated by the 2-3
mixing in the neutrino sector, and hence both the large and small
mixing cases give almost the same $\Delta^{(R)}_{23}$.  (Therefore, we
plot only $\Delta^{(R)}_{23}$ in the large mixing case.)  On the
contrary, with the choice of $V_L$ given in Eqs.~(\ref{VL_la}) and
(\ref{VL_sa}), $[V_L]_{31}=0$, and hence both $\Delta^{(R)}_{12}$ and
$\Delta^{(R)}_{13}$ are from the term proportional to $y_{\nu_2}^2$ in
Eq.~(\ref{m_dR}).  As a result, these quantities are more enhanced for
the large angle MSW case because of the larger $|[V_L]_{21}|$.
Furthermore, since $|[V_L]_{22}|\simeq |[V_L]_{23}|$ in the mixing
matrices we use, $|\Delta^{(R)}_{12}|\simeq |\Delta^{(R)}_{13}|$ is
realized both in the large and small mixing cases.  If $[V_L]_{31}\sim
O(10^{-2})$, however, $y_{\nu_3}^2$-term may give comparable
contributions.  We also found that $|\Delta^{(R)}_{ij}|$ is almost
independent of the phases $\phi^{(Q)}$ and $\phi^{(L)}$.  As suggested
from Eq.~(\ref{m_dR}), however, the phase of $\Delta^{(R)}_{ij}$ is
sensitive to $\phi^{(L)}$; the phase of $\Delta^{(R)}_{ij}$ agrees
with $\phi^{(L)}_i-\phi^{(L)}_j$ very accurately.  Therefore,
$\Delta^{(R)}_{ij}$ may have an arbitrary phase and can be a new
source of CP violation.

The most stringent constraints on the off-diagonal elements in the
squark mass matrices are from the $\Delta S=2$ and $\Delta B=2$
processes \cite{NPB477-321}, so we discuss them in this paper.  In
particular, $|\Delta^{(R)}_{12}|\sim O(10^{-4})$ realized with
$M_{\nu_R}\sim 10^{14}-10^{15}~{\rm GeV}$ may generate the
$\epsilon_K$ parameter as large as the experimental value
\cite{NPB477-321}.  In order to discuss the SUSY contribution to the
flavor and CP violations (in particular, to the $\epsilon_K$
parameter), we calculate the effective Hamiltonian contributing to
$\Delta S=2$ and $\Delta B=2$ processes:\footnote{In Eq.~(\ref{Heff}),
possible operators arising from the left-right squark mixing are
omitted, although those are included in our numerical calculation.  We
checked their contributions are negligible relative to the dominant
contributions from $\Delta^{(R)}_{ij}$ (and $\Delta^{(L)}_{ij}$).}
\begin{eqnarray}
{\cal H}_{\rm eff} &=& 
\left[ C_{LL} \right]_{ij}
\left( \bar{d}_i^a \gamma_\mu P_L d_j^a \right)
\left( \bar{d}_i^b \gamma_\mu P_L d_j^b \right)
+ \left[ C_{RR} \right]_{ij}
\left( \bar{d}_i^a \gamma_\mu P_R d_j^a \right)
\left( \bar{d}_i^b \gamma_\mu P_R d_j^b \right)
\nonumber \\ &&
+ \left[ C^{(1)}_{LR} \right]_{ij}
\left( \bar{d}_i^a P_L d_j^a \right)
\left( \bar{d}_i^b P_R d_j^b \right)
+ \left[ C^{(2)}_{LR} \right]_{ij}
\left( \bar{d}_i^a P_L d_j^b \right)
\left( \bar{d}_i^b P_R d_j^a \right),
\label{Heff}
\end{eqnarray}
where $P_{R,L}=\frac{1}{2}(1\pm\gamma_5)$, $d_i^a$ is the down-type
quark in $i$-th generation with $a$ being SU(3)$_C$ index.  Based on
the operator structure, we call the first and second operators $LL$
and $RR$ operators, respectively, while the third and fourth ones $LR$
operators.  In our analysis, we only include the dominant
contributions from the gluino-squark loops, and use mass-insertion
approximation.  The SUSY contributions to the coefficients are found
in Ref.~\cite{NPB477-321}, and are given by
\begin{eqnarray}
\left[ C_{LL} \right]_{ij} &=& 
- \frac{\alpha_s}{36m_{\tilde{q}}^2}
\left( \Delta^{(L)}_{ij} \right)^2 
\left[ 4xf_6(x) + 11\tilde{f}_6(x) \right],
\\
\left[ C_{RR} \right]_{ij} &=& 
- \frac{\alpha_s}{36m_{\tilde{q}}^2}
\left( \Delta^{(R)}_{ij} \right)^2 
\left[ 4xf_6(x) + 11\tilde{f}_6(x) \right],
\\
\left[ C^{(1)}_{LR} \right]_{ij} &=& 
- \frac{\alpha_s}{3m_{\tilde{q}}^2}
\Delta^{(L)}_{ij} \Delta^{(R)}_{ij}
\left[ 7xf_6(x) - \tilde{f}_6(x) \right],
\\
\left[ C^{(2)}_{LR} \right]_{ij} &=& 
- \frac{\alpha_s}{9m_{\tilde{q}}^2}
\Delta^{(L)}_{ij} \Delta^{(R)}_{ij}
\left[ xf_6(x) + 5\tilde{f}_6(x) \right],
\label{C_LR}
\end{eqnarray}
where $x=m_{\tilde{G}}^2/m_{\tilde{q}}^2$, and 
\begin{eqnarray}
f_6(x) &=& \frac{6(1+3x)\log x + x^3 - 9x^2 - 9x + 17}{6(x-1)^5},
\\
\tilde{f}_6(x) &=& 
\frac{6x(1+x)\log x - x^3 - 9x^2 + 9x + 1}{6(x-1)^5}.
\end{eqnarray}
Furthermore, $\Delta^{(L)}_{ij}$ parameterizes the flavor violation in
the left-handed down-type squarks $\tilde{d}_L$:
\begin{eqnarray}
\Delta^{(L)}_{ij}\equiv 
\frac{[m^2_{\tilde{d}_L}]_{ji}}{m^2_{\tilde{q}}}.
\end{eqnarray}
With the minimal supergravity boundary condition,
$[m^2_{\tilde{d}_L}]_{ij}$ is generated by the large top Yukawa
coupling $y_t$ and the CKM matrix.  Approximately,
\begin{eqnarray}
[m^2_{\tilde{d}_L}]_{ij} \simeq
-\frac{1}{8\pi^2} 
y_t^2 \left[ V_{\rm CKM}^* \right]_{3i} 
\left[ V_{\rm CKM} \right]_{3j} (3m_0^2+a_0^2)
\left( 3 \log \frac{M_*}{M_{\rm GUT}}
+ \log \frac{M_{\rm GUT}}{M_{\rm weak}} \right),
\label{m_dL}
\end{eqnarray}
and numerically, for example, $|\Delta^{(R)}_{12}|$ is $O(10^{-4})$.

With the effective Hamiltonian (\ref{Heff}), we first calculate the
SUSY contribution to $\epsilon_K$:
\begin{eqnarray}
\epsilon_K = 
\frac
{e^{i\pi/4} {\rm Im} \langle K| {\cal H}_{\rm eff} |\bar{K}\rangle}
{2\sqrt{2}m_K \Delta m_K},
\end{eqnarray}
where $m_K$ and $\Delta m_K$ are the $K$-meson mass and the
$K_L$-$K_S$ mass difference, respectively.  QCD corrections below the
electroweak scale are neglected.  The matrix element of the effective
Hamiltonian is calculated by using the vacuum insertion approximation:
\begin{eqnarray}
\langle K | {\cal H}_{\rm eff} | \bar{K}\rangle &=&
\frac{2}{3} \left( \left[ C_{LL} \right]_{12}
+ \left[ C_{RR} \right]_{12} \right) m_K^2 f_K^2
\nonumber \\ &&
+ \left[ C^{(1)}_{LR} \right]_{12}
\left( \frac{1}{2} \frac{m_K^2}{(m_s+m_d)^2} 
+  \frac{1}{12} \right) m_K^2 f_K^2
\nonumber \\ &&
+ \left[ C^{(2)}_{LR} \right]_{12}
\left( \frac{1}{6} \frac{m_K^2}{(m_s+m_d)^2} 
+ \frac{1}{4} \right) m_K^2 f_K^2,
\label{<H>}
\end{eqnarray}
where $f_K\simeq 160~{\rm MeV}$ is the $K$-meson decay constant, and
$m_s$ and $m_d$ are the strange and down quark masses, respectively.

The new flavor and CP violating effects in the $\tilde{d}_R$ sector
have an important implication to the calculation of $\epsilon_K$,
since the matrix element of the $LR$-type operators are enhanced by
the factor $(m_K/m_s)^2\simeq 10$, as seen in Eq.~(\ref{<H>}).  As a
result, if all the coefficients of the $LL$, $RR$, and $LR$ operators
are comparable, the $LR$ ones dominate the contribution to the
$\epsilon_K$ parameter.  In the SUSY SU(5) model with the right-handed
neutrinos, this is the case.  However, if there is no right-handed
neutrinos, or if the running above the GUT scale is not included,
$\Delta^{(R)}_{12}$ is much more suppressed and the $LL$ operator
gives the dominant effect.  Then, the SUSY contribution to
$\epsilon_K$ becomes much smaller \cite{PRD53-5233, hph9908449}.

\begin{figure}
\centerline{\epsfxsize=0.5\textwidth\epsfbox{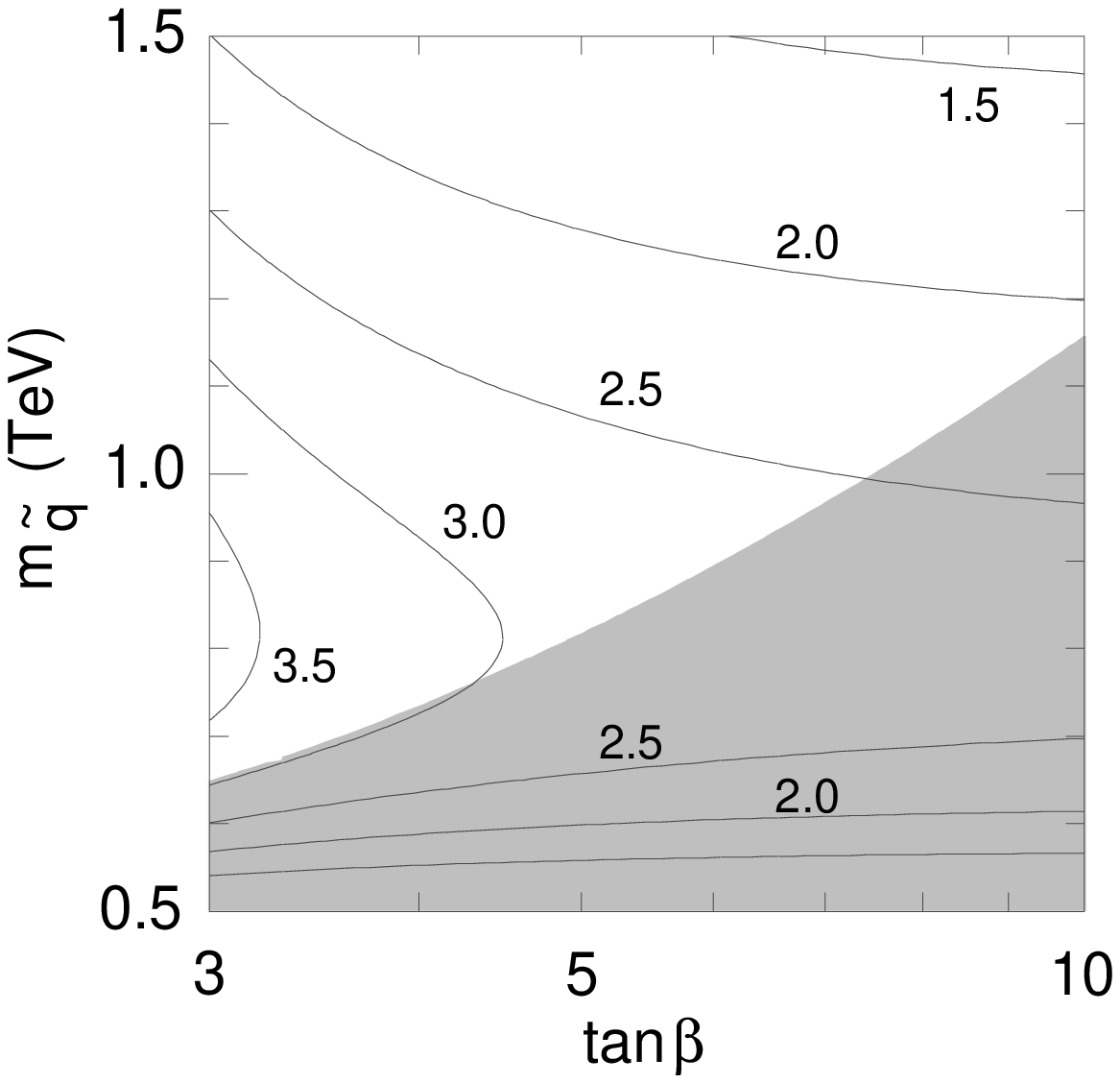}}
\caption{Contours of constant $|\epsilon_K^{\rm (SUSY)}|$ (in units of
$10^{-3}$) on $\tan\beta$ vs.\ $m_{\tilde{q}}$ plane for the large
angle MSW case (\ref{VL_la}).  We take $m_{\tilde{W}}=150~{\rm GeV}$,
$M_{\nu_R}=2\times 10^{14}~{\rm GeV}$, $a_0=0$, and the GUT phases are
chosen so that $|\epsilon_K^{\rm (SUSY)}|$ is maximized.  The shaded
region corresponds to $Br(\mu\rightarrow e\gamma)\ge 4.9\times
10^{-11}$.}
\label{fig:epsla}
\vskip .5in
\centerline{\epsfxsize=0.5\textwidth\epsfbox{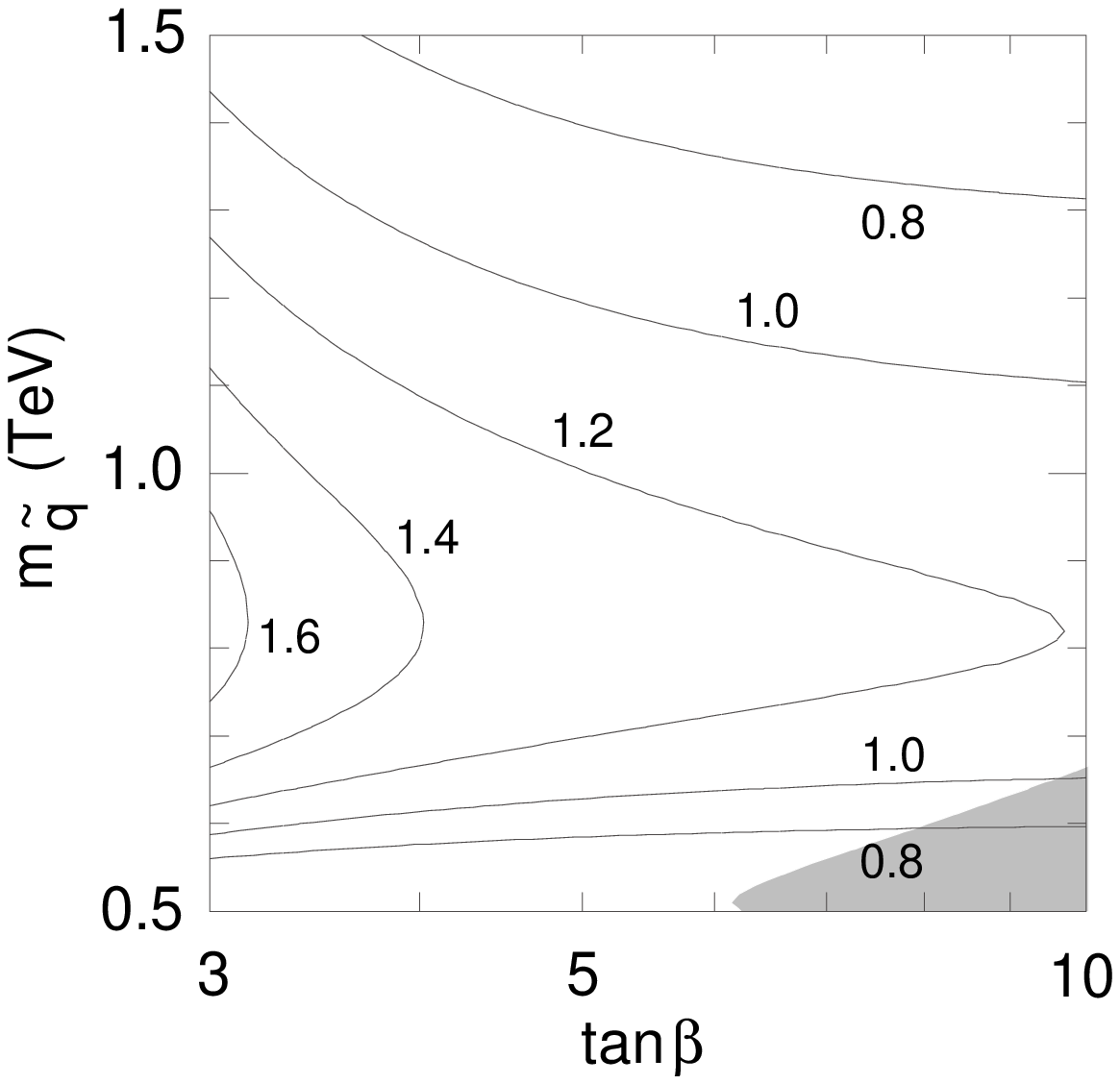}}
\caption{Same as Fig.~\ref{fig:epsla} except the neutrino mass matrix
suggested by the small angle MSW case (\ref{VL_sa}), and
$M_{\nu_R}=10^{15}~{\rm GeV}$.}
\label{fig:epssa}
\end{figure}

We solve the RG equation numerically and obtain the electroweak scale
values of the MSSM parameters.  Then, we calculate $\epsilon_K^{\rm
(SUSY)}$, the SUSY contribution to the $\epsilon_K$ parameter.  In
Figs.~\ref{fig:epsla} and \ref{fig:epssa}, we show $|\epsilon_K^{\rm
(SUSY)}|$ on $\tan\beta$ vs.\ $m_{\tilde{q}}$ plane.  Here, we take
$m_{\tilde{W}}=150~{\rm GeV}$, and consider both the large and small
angle MSW cases.  The result depends on the phase
$\phi^{(L)}_2-\phi^{(L)}_1$; $\epsilon_K^{\rm (SUSY)}$ is
approximately proportional to $\sin (\phi^{(L)}_2-\phi^{(L)}_1+{\rm
arg}(\Delta^{(L)}_{12}))$.  We choose $\phi^{(L)}_2-\phi^{(L)}_1$
which maximizes $|\epsilon_K^{\rm (SUSY)}|$.  The RG effect also
generates off-diagonal elements in the slepton mass matrix, which
induce various lepton flavor violations.  In particular, in our case,
rate for the $\mu\rightarrow e\gamma$ process may become significantly
large.  In Figs.~\ref{fig:epsla} and \ref{fig:epssa}, we shaded the
region with $Br(\mu\rightarrow e\gamma)\ge 4.9\times 10^{-11}$, which
is already excluded by the negative search for $\mu\rightarrow
e\gamma$ \cite{PDG}.  The $\mu\rightarrow e\gamma$ process is enhanced
for large $\tan\beta$ \cite{PLB357-579, PRD53-413, PRD53-2442}.
However, large amount of parameter space is still allowed, in
particular, in the low $\tan\beta$ region.  As one can see, the SUSY
contribution to $\epsilon_K$ can be $O(10~\%)$ of the experimentally
measured value ($\epsilon_K^{(\rm exp)}\simeq 2.3\times 10^{-3}$
\cite{PDG}) even with relatively heavy squarks ($m_{\tilde{q}}\sim
1~{\rm TeV}$), and can be even comparable to $\epsilon_K^{(\rm exp)}$.
Of course, if $M_{\nu_R}$ is smaller, or if there is an accidental
cancellation among the phases, then the SUSY contribution to
$\epsilon_K$ is more suppressed.

The information from $\epsilon_K$ plays a significant role in
constraining the allowed region in the so-called $\rho$ vs.\ $\eta$
plane \cite{hph9908520}.  If the currently measured $\epsilon_K$ has a
significant contamination of the SUSY contribution, however, $\rho$
and $\eta$ suggested by $\epsilon_K$ may become inconsistent with
those from other constraints.  Most importantly, the $B$ factories
will measure the CP violation in the $B_d\rightarrow \psi K_S$ mode,
and in the standard model, this process gives the phase ${\rm
arg}(-[V_{\rm CKM}]_{cd}[V_{\rm CKM}^*]_{cb}/ [V_{\rm
CKM}]_{td}[V_{\rm CKM}^*]_{tb})$.  As will be discussed below, the
SUSY correction to this process is small at least in the parameter
space we are discussing.  By comparing $\rho$ and $\eta$ suggested by
$\epsilon_K$ and those from $B_d\rightarrow \psi K_S$, we may see an
anomaly arising from the SUSY loop effect if we fit the data just
assuming the standard model.

Other check point is the $K_L$-$K_S$ mass difference
\begin{eqnarray}
\Delta m_K = 
\frac{|\langle K|{\cal H}_{\rm eff}|\bar{K}\rangle |}{m_K},
\end{eqnarray}
since, in some case, SUSY contribution to $\Delta m_K$ is
significantly large.  In our case, however, $\Delta m_K$ is less
important than the $\epsilon_K$ parameter; with
$\Delta^{(R)}_{12}\sim\Delta^{(L)}_{12}\sim O(10^{-4}-10^{-3})$, the
SUSY contribution to the $\Delta m_K$ parameter is $O(1~\%)$ level of
its experimental value.

Finally, we discuss the contributions to $\Delta B=2$ processes.
Since the $\Delta^{(R,L)}_{13}$ and $\Delta^{(R,L)}_{23}$ parameters
are non-vanishing, they may change the standard model predictions to
the $B$ decays.  In particular, due to the possible GUT phases
$\phi^{(Q)}$ and $\phi^{(L)}$, the CP violations in $B$ systems as
well as the mass differences $\Delta m_B$ may be affected.
$\Delta^{(R,L)}_{13}$ affects the $B_d$ meson.  With
$\Delta^{(R)}_{13}\lesssim O(10^{-3})$ and $\Delta^{(L)}_{13}\sim
O(10^{-4})$, however, SUSY contribution to $\Delta m_{B_d}$ is also
small, and is at most $\sim 1~\%$ \cite{NPB353-591}.  This also means
that the phase in the $B_d$-$\bar{B}_d$ mixing matrix is dominated by
the standard model contribution.

Because of the large 2-3 mixing in the $V_L$ matrix, the SUSY
contribution to the $\Delta m_{B_s}$ parameter is more enhanced.  In
the standard model, the ratio $\Delta m_{B_s}/\Delta m_{B_d}$ is
approximately $|[V_{\rm CKM}]_{ts}/[V_{\rm CKM}]_{td}|^2$.  With this
relation, we found that the SUSY contribution to $\Delta m_{B_s}$ can
be as large as $\sim 10~\%$ of the standard model contribution when
$y_{\nu_3}$ is maximally large.  This fact also suggests that a
sizable correction may be possible to the phase in the
$B_s$-$\bar{B}_s$ mixing matrix because $\Delta^{(R)}_{23}$ has a
phase equal to $e^{-i(\phi^{(L)}_2-\phi^{(L)}_3)}$.  Very small CP
asymmetry is expected in the decay modes like
$B_s\rightarrow\phi\eta$, $D_s\bar{D}_s$, and $\phi\eta'$ in the
standard model, and the new source of the CP violation may affect the
standard model predictions.  Notice that, in the model without the
right-handed neutrinos, $\Delta^{(L)}_{23}$ is approximately
proportional to $[V_{\rm CKM}^*]_{ts} [V_{\rm CKM}]_{tb}$, while
$\Delta^{(R)}_{23}$ is negligibly small.  Then, the SUSY contribution
to the matrix element $\langle B_s|{\cal H}_{\rm
eff}|\bar{B}_s\rangle$ has almost the same phase as the standard model
contribution, and hence the phase in the $B_s$-$\bar{B}_s$ mixing
matrix is not significantly affected.

In summary, in the SUSY SU(5) model with the right-handed neutrinos,
sizable CP and flavor violations are possible in the $\tilde{d}_R$
sector, which is not the case in models without the right-handed
neutrinos.  Importantly, with such effects, the SUSY contribution to
the $\epsilon_K$ parameter can be as large as the experimental value
($\epsilon_K^{(\rm exp)}\simeq 2.3\times 10^{-3}$) if the right-handed
neutrino mass $M_{\nu_R}$ is high enough.  In this paper, we only
considered the $\Delta S=2$ and $\Delta B=2$ processes.  In the
future, however, other CP violation informations will be available, in
particular, in $\Delta S=1$ and $\Delta B=1$ processes.  Given the
fact that there can be extra sources of the CP violation, it would be
desirable to measure as much CP violations as possible and to test
standard model predictions.

{\sl Note added}: After the completion of the main part of this work,
the author was noticed a paper by S.\ Baek, T.\ Goto, Y.\ Okada and
K.\ Okumura \cite{hph0002142} which discussed implications of the
right-handed neutrinos to the $B$ physics in the SUSY GUT, in
particular, to the $\Delta m_{B_s}$ parameter.  In their paper,
however, effects of the GUT phases $\phi^{(Q)}$ and $\phi^{(L)}$ are
not discussed.

{\sl Acknowledgments}: The author would like to thank J.\ Hisano for
useful discussions.  This work was supported by the National Science
Foundation under grant PHY-9513835, and also by the Marvin L.\
Goldberger Membership.


\newpage

\begin{thebibliography}{99}

\bibitem{SuzTot}
Y.\ Suzuki and Y.\ Totsuka (eds.),
  {\sl Neutrino Physics and Astrophysics} (Elsevier Science, 
  Amsterdam, 1999).

\bibitem{Yanagida}
T.\ Yanagida, in {\sl Proceedings of the Workshop on Unified Theory and
  Baryon Number of the Universe}, eds.\ O.\ Sawada and A.\ Sugamoto (KEK,
  1979) p.95.

\bibitem{GelRamSla}
M.\ Gell-Mann, P.\ Ramond and R.\ Slansky, in {\sl Supergravity},
  eds.\ P.\ van Niewwenhuizen and D.\ Freedman (North Holland,
  Amsterdam, 1979).

\bibitem{PRD44-817}
P.\ Langacker and M.\ Luo,
  {\sl Phys.\ Rev.} {\bf D44} (1991) 817.

\bibitem{PLB260-447}
U.\ Amaldi, W.\ de Boer and H.\ F\"urstenau,
  {\sl Phys.\ Lett.} {\bf B260} (1991) 447.

\bibitem{PRL88-1562}
Super-Kamiokande Collaboration (Y. Fukuda {\sl et al.}),
  {\sl Phys.\ Rev.\ Lett.} {\bf 81} (1998) 1562.

\bibitem{NPB449-437}
R.\ Barbieri, L.\ Hall and A.\ Strumia,
  {\sl Nucl.\ Phys.} {\bf B449} (1995) 437.

\bibitem{PDG}
Particle Data Group (C.\ Caso {\sl et al.}),
  {\sl Eur.\ Phys.\ J.} {\bf C3} (1998) 1.

\bibitem{PRD53-413}
N.\ Arkani-Hamed, H.C.\ Cheng and L.J.\ Hall,
  {\sl Phys.\ Rev.} {\bf D53} (1996) 413.

\bibitem{PRD58-116010}
J.\ Hisano, D.\ Nomura, Y.\ Okada, Y.\ Shimizu and M.\ Tanaka,
  {\sl Phys.\ Rev.} {\bf D58} (1998) 116010.

\bibitem{NPB267-415}
L.J. Hall, V.A. Kostelecky and S. Raby, 
  {\sl Nucl.\ Phys.} {\bf B267} (1986) 415.

\bibitem{PRL57-961}
F.~Borzumati and A.~Masiero, 
  {\sl Phys.\ Rev.\ Lett.} {\bf 57} (1986) 961.

\bibitem{PLB321-56}
T.\ Moroi,
  {\sl Phys.\ Lett.} {\bf B321} (1994) 56.

\bibitem{PLB338-212}
R.~Barbieri and L.J.~Hall,
  {\sl Phys.\ Lett.} {\bf B338} (1994) 212.

\bibitem{NPB445-219}
R.\ Barbieri, L.J.\ Hall and A.\ Strumia,
  {\sl Nucl.\ Phys.} {\bf B445} (1995) 219.

\bibitem{PLB391-341}
J.\ Hisano, T.\ Moroi, K.\ Tobe and M.\ Yamaguchi,
 {\sl Phys.\ Lett.} {\bf B391} (1997) 341.

\bibitem{NPB458-3}
P.\ Ciafaloni, A.\ Romanino and A.\ Strumia,
  {\sl Nucl.\ Phys.} {\bf B458} (1996) 3.

\bibitem{PLB357-579}
J.\ Hisano, T.\ Moroi, K.\ Tobe, M.\ Yamaguchi and T.\ Yanagida,
  {\sl Phys.\ Lett.} {\bf B357} (1995) 579.

\bibitem{PRD53-2442}
J.\ Hisano, T.\ Moroi, K.\ Tobe and M.\ Yamaguchi,
  {\sl Phys.\ Rev.} {\bf D53} (1996) 2442.

\bibitem{PRD59-116005}
J.\ Hisano and D.\ Nomura,
  {\sl Phys.\ Rev.} {\bf D59} (1999) 116005.

\bibitem{PRD39-3447}
T.\ Kurimoto,
  {\sl Phys.\ Rev.} {\bf D39} (1989) 3447.

\bibitem{NPB353-591}
S.\ Bertolini, F.\ Borzumati, A.\ Masiero and G.\ Ridolfi,
  {\sl Nucl.\ Phys.} {\bf B353} (1991) 591.

\bibitem{PRD53-5233}
T.\ Goto, Y.\ Okada and T.\ Nihei,
  {\sl Phys.\ Rev.} {\bf D53} (1996) 5233.

\bibitem{hph9908449}
T.\ Goto, Y.\ Okada and Y.\ Shimizu,
  hep-ph/9908449.

\bibitem{hph9802201} 
S.M.\ Bilenky and C.\ Giunti, hep-ph/9802201.

\bibitem{PLB420-397}
CHOOZ Collaboration (M.\ Apollonio {\sl et al.}),
  {\sl Phys.\ Lett.} {\bf B420} (1998) 397.

\bibitem{NPB477-321}
F.\ Gabbiani, E.\ Gabrielli, A.\ Masiero and L.\ Silvestrini,
  {\sl Nucl.\ Phys.} {\bf B477} (1996) 321.

\bibitem{hph9908520}
See, for example, A.F.\ Falk, hep-ph/9908520.

\bibitem{hph0002142}
S.\ Baek, T.\ Goto, Y.\ Okada and K.\ Okumura,
  hep-ph/0002141.

\end{thebibliography}
\end{document}